\documentstyle[preprint,aps]{revtex}
\preprint{USM-TH-94}
\begin{document}
\title{Note on the dual BRST Symmetry in U(1) Gauge Theory}
\author{ Patricio Gaete \thanks{E-mail: pgaete@fis.utfsm.cl}}
\address{Departamento de F\'{\i}sica, Universidad T\'ecnica F.
Santa Maria, Valpara\'{\i}so, Chile} \maketitle

\begin{abstract}
We analyze the relation between the Lagrangian and Hamiltonian
BRST symmetry generators for a recently proposed two-dimensional
symmetry. In particular it is shown that this symmetry may be
obtained from a canonical transformation in the ghost sector in
a gauge independent way.
\end{abstract}
\smallskip

PACS number(s): 12.20.-m, 11.30.Ly

\section{INTRODUCTION}

Nowadays the concept of BRST symmetry\cite{Becchi} plays an
essential role in the quantization of gauge theories. As is well
known the BRST formalism has been very useful in the framework of
path integral quantization, where the BRST generator (charge) is a
key ingredient of the effective action, as well as it finds
interesting applications in the operator formulation of the
theory. An illustrative example on this subject arises when one
considers string theory. In fact, following the BRST inspired
approach it was possible to derive the string critical dimensions
in a straightforward and economical way\cite{Claudio}. In this
context it may be recalled that there exists two approaches to the
BRST formulation of gauge theories. One is based on the
Hamiltonian formulation, where the BRST charge is constructed in
terms of the constraints and the higher order structure functions
in a gauge independent way. The other approach is based on the
Lagrangian formulation, in such a case the BRST charge is computed
from a gauge-fixed Lagrangian by using Noether's prescription. In
passing we also recall that in the path integral quantization
formalism (both Lagrangian and Hamiltonian) the original gauge
invariance is incorporated by means of the extension of the phase
space including ghost fields. Thus the main idea is to substitute
the local gauge invariance by a rigid Grassmannian symmetry (or
global supersymmetry) known as the BRST symmetry. In this way one
assigns a global nilpotent charge to this symmetry whose
cohomology produces the physical states.

On the other hand, recently a great deal of attention has been
devoted to the study of new symmetries in gauge theories. For
instance, Lavelle and McMullan\cite{Lavelle} found that QED
displays a new nonlocal and noncovariant symmetry. In such a case
the symmetry transformations are compatible with the gauge-fixing
conditions. At the same time Tang and Finkelstein\cite{Tang}
constructed a nonlocal but covariant symmetry for QED. Let us also
mention here that Yang and Lee\cite{Yang} derived a noncovariant
but nonlocal symmetry of QED. More recently, Malik\cite{Malik}
showed that in two dimensions of spacetime there exists a local,
covariant and nilpotent BRST symmetry, the so-called dual
symmetry, under which the gauge-fixing term remains invariant for
a free U(1) gauge theory and QED. Furthermore, this author claimed
that this symmetry transformation is not the generalization of the
above symmetries in two dimensions of spacetime. It is worth
stressing at this stage that despite their relevance these studies
have been, however, carried out in the gauge fixed scheme only.

Meanwhile, in a previous paper\cite{Gaete} we have discussed the
relation between the Lagrangian and Hamiltonian symmetry
generators for the Lavelle and McMullan's symmetry in a gauge
invariant way using the Batalin-Fradkin-Vilkovisky formalism. In
particular, we have showed that the Lavelle and McMullan's
symmetry may be derived from a canonical transformation in the
ghost sector. We also recall that there are definite advantages of
the Hamiltonian approach over the conventional gauge fixed
analysis. Ambiguities related to gauge fixing conditions are
avoided and it does not need an auxiliary field to construct an
off-shell nilpotent symmetry transformation. Let us also mention
here that a similar analysis has been made, independently, by
Rivelles\cite{Rivelles}. We are thus motivated to investigate in
this paper whether the so-called dual symmetry is a new symmetry
or it is merely an artifact of the canonical transformation in the
ghost sector.

The outline of this paper is as follows. In Sect.2 we briefly
recap the BFV-BRST formalism for a free U(1) gauge theory in four
dimensions of spacetime. This will form the basis of our
subsequent considerations. In Sect.3 we will focus our attention
to the two-dimensional case. Particular care is paid to establish
a direct connection between the Lagrangian and Hamiltonian BRST
symmetry generators for the so-called dual symmetry.

\section{General Considerations on the BFV-BRST Formalism}

Let us commence our considerations with a short presentation of
the BFV-BRST formalism for a free U(1) gauge theory in four
dimensions. It should be noted that this method is a general
procedure for quantizing systems with first class constraints. A
detailed discussion of the formalism can be found in
\cite{Claudio}. We summarize the essence of this formalism in
terms of a finite number of phase-space variables, this makes the
discussion simpler. In such a case the action for the theory under
consideration is taken to be
\begin{equation}
S = \int {dt} \left( {p^\mu  \stackrel {\bf{\cdot}}{q_\mu }  - H_0
- \lambda ^a \varphi _a } \right), \label{bfv1}
\end{equation}
where the coordinates $(q_\mu,p^\mu)$ are the canonical variables
describing the theory. The canonical Hamiltonian is $H_0$, and
$\lambda ^a$ are the Lagrange multipliers associated with the
first class constraints $\varphi _a$. As prescribed by the general
theory the Lagrange multipliers are treated in the same foot as
the canonical variables, thus we introduce conjugate canonical
momenta to $\lambda ^a$, say $p_a$. Evidently, the ${p_a}$'s must
be imposed as new constraints in order that the dynamics of the
theory does not change. Now, the BFV approach introduces a pair of
canonically conjugate ghosts $(C^a(x),{\cal P}^a(x))$ for each
constraints. The Poisson algebra of these ghosts is
\begin{equation}
\left[ {C\left( {{\bf x},t} \right), {\cal P}\left( {{\bf y},t}
\right)} \right] = - \delta \left( {{\bf x} - {\bf y}} \right),
\label{bfv2}
\end{equation}
where $C$ and $\cal P$ has ghost number $1$ and $-1$,
respectively. These considerations naturally lead to an extended
phase space, where we have substituted the local gauge invariance
by a global supersymmetry invariance (BRST invariance). In this
extended phase space the generator of the BRST symmetry for a
theory with first class constraints has the form
\begin{equation}
\Omega  = C_a \varphi ^a  + \frac{1}{2}P^a f_a^{bc} C_b C_c +...,
\label{bfv3}
\end{equation}
where $f_a^{bc}$ are the structure functions, and $\Omega$ is by
construction nilpotent ($ \left[ \Omega,\Omega \right]=0 $). We
also recall that, at the quantum level, in the extended phase
space there exists the Fradkin-Vilkovisky theorem
\cite{Claudio,Fradkin}. This theorem states that the functional
integral
\begin{equation}
{\cal Z}_\Psi   = \int {{\cal D}\mu } \exp \left( {iS_{eff} }
\right), \label{bfv4}
\end{equation}
where the effective action $S_{eff}$ is given by
\begin{equation}
S_{eff}  = \int {dt\left( {p^\mu  \stackrel {\bf{\cdot}}{q_\mu } +
C^a \stackrel {\bf{\cdot}}{\cal P}_a  + p^a \stackrel
{\bf{\cdot}}{\lambda} _a - H_0  - \left[ {\Omega ,\Psi } \right]}
\right)}, \label{bfv5}
\end{equation}
being independent of the choice of $\Psi$. Here $\Psi$ is an
arbitrary fermionic gauge-fixing function, and ${\cal D}\mu$ is
the Liouville measure on the phase space. This concludes our brief
review of the BFV formalism.

Let us now proceed to apply the above procedure for a free U(1)
gauge theory, in other words,
\begin{equation}
{\cal L} =  - \frac{1}{4}F_{\mu \nu } F^{\mu \nu }. \label{bfv6}
\end{equation}
Then, from (\ref{bfv1}) the canonical action takes the form
\begin{equation}
S = \int {dx} \left( {\stackrel {\bf{\cdot}}{A_i} \Pi ^i  - H_0  -
\lambda \varphi } \right), \label{bfv7}
\end{equation}
where $\Pi ^i$ is the momenta conjugate to $A_i$. $H_0$ is the
canonical Hamiltonian, that is,
\begin{equation}
H_0  = \int {d^3 } x\left( { - \frac{1}{2}\Pi _i \Pi ^i  +
\frac{1}{4}F_{ij} F^{ij}  + \Pi _i \partial ^i A_0 } \right),
\label{bfv8}
\end{equation}
and it is straightforward to see that the preservation in time of
the constraint primary $(\Pi^0=0)$ leads to the secondary
constraint
\begin{equation}
\varphi  = \partial _i \Pi ^i  = 0. \label{gauss}
\end{equation}
We mention in passing that in the action (\ref{bfv7}) the
canonical variables $A_0$ and $ \Pi^0$ have been omitted because $
\Pi^0=0 $, which does not represent a true dynamical degree of
freedom of the theory. Thus, $A_0$ can be absorbed by redefining
the multiplier $\lambda$, i. e., $\lambda$ and $A_0$ do not need
to be treated as independent variables. With this at hand, the
effective action then reads:
\begin{equation}
S_{eff}  = \int {d^4 x\left( {\Pi ^i  \stackrel {\bf{\cdot}}{A_i }
+ {\cal P}\stackrel {\bf{\cdot}}{C}+{\overline {\cal P}} \stackrel
{\bf{\cdot}}{\overline C} + \Pi^0 \stackrel {\bf{\cdot}}{A_0} -
H_0 - \left[ {\Omega ,\Psi } \right]} \right)}, \label {bfv9}
\end{equation}
where we have introduced the antighost pair $({\overline
C}(x),{\overline {\cal P}}(x))$  with respective ghost numbers
$-1$ and $+1$, and satisfying the Poisson algebra (\ref{bfv2}).
The BRST charge $\Omega$ can be easily given as
\begin{equation}
\Omega  = \int {d^3 } x\left( {C\partial _i \Pi ^i  - i{\overline
{\cal P}}\Pi _0} \right). \label {bfv1000}
\end{equation}
We can now write the corresponding transformations generated by
$\Omega$, that is,
\begin{equation}
\delta A_i  = - \varepsilon \partial _i C , \label{bfv10}
\end{equation}
\begin{equation}
\delta A_0  = -i\varepsilon {\overline {\cal P}} , \label{bfv11}
\end{equation}
\begin{equation}
\delta \Pi_i = 0 , \label{bfv12}
\end{equation}
\begin{equation}
\delta \Pi _0 = 0 , \label{bfv13}
\end{equation}
\begin{equation}
\delta C = 0 , \label{bfv13}
\end{equation}
\begin{equation}
\delta {\overline C} =  i\varepsilon  \Pi _0, \label{bfv14}
\end{equation}
\begin{equation}
\delta {\overline {\cal P}} = 0, \label{bfv15}
\end{equation}
\begin{equation}
\delta {\cal P} = -\varepsilon \partial _i {\Pi ^i}, \label{bfv16}
\end{equation}
where $\varepsilon$ is an anticommuting spacetime independent
infinitesimal parameter. In order to compute the effective action
(\ref{bfv9}), we have to select the gauge fixing function $\Psi $.
There are a variety of these which have been found useful and
convenient in different calculational context. We can choose, for
example, $\Psi$ in the form
\begin{equation}
\Psi  = \int {d^3 } x\left( {{\cal P}A_0  - i{\overline C}\left(
{\frac{{x_i A^i }}{{x^2 }} - \frac{\xi }{2}\Pi ^0  - \stackrel
{\bf{\cdot}}A_0 } \right)} \right), \label{FockS}
\end{equation}
which leads to the modified Fock-Schwinger gauge \cite{Gaete2}.
However, of this turn, we take $\Psi$ as
\begin{equation}
\Psi  = \int {d^3 x} \left( {{\cal P}A_0  - i{\overline C}\left(
{\partial _i \Pi ^i - \frac{\xi }{2}\Pi _0 } \right)} \right),
\label{bfv17}
\end{equation}
where $\xi$ is a real parameter that describes a set of gauges.
Explicitly, for  $\xi = 0,1$ and infinity  we obtain the Landau,
Feynman and unitary gauges, respectively. Plugging this expression
into (\ref{bfv9}), we find that the resulting effective action is
given by
\begin{equation}
S_{eff}  = \int {d^4 } x\left( { - \frac{1}{4}F^{\mu \nu } F_{\mu
\nu }  + i\overline C \partial _\mu  \partial ^\mu  C +
\frac{1}{{2\xi }}\left( {\partial _\mu  A^\mu  } \right)^2 }
\right).\label{bfv18}
\end{equation}
We immediately recognize the above to be the same as the
Lagrangian effective action.

Before concluding this section we call attention to the fact that
in contrast to the gauge-fixing term, the gauge field $F_{\mu\nu}$
remains invariant under the transformation generated by $\Omega$
(\ref{bfv1000}), that is, $\delta F_{\mu\nu}=0$ and $\delta \left(
{\partial _\mu  A^\mu  } \right) =  - \varepsilon \left( {i\mathop
{\overline {\cal P} }\limits^ \cdot   - \nabla ^2 C} \right)$.
However, it is possible to recast the BRST charge (\ref{bfv1000})
which corresponds to a nilpotent symmetry transformation under
which the gauge-fixing term remains invariant.  This can be done
by a suitable canonical transformation in the BFV phase space, in
such a way that any two BRST generators are related by such
transformations \cite{Claudio}. In the work of Ref.\cite{Gaete},
we had showed that by performing the following canonical
transformation in the ghost sector:
\begin{equation}
C^\prime   =   \frac{1}{\nabla^2 }{\cal P}, \label{canno1}
\end{equation}
\begin{equation}
{\cal P}^\prime = {\nabla^2 }C, \label{canno2}
\end{equation}
\begin{equation}
{\overline C}^\prime  =  -  \overline {\cal P}, \label{canno3}
\end{equation}
\begin{equation}
{\overline {\cal P}}^\prime  = - \overline C, \label{canno4}
\end{equation}
the gauge-fixing term remains invariant. In effect, as a
consequence of this canonical transformation, the new BRST charge
$\Omega^{\bot}$ then becomes
\begin{equation}
\Omega ^ \bot   = \int {d^3 } x\left( {\frac{1}{{\nabla ^2 }}{\cal
P}\partial _i \Pi ^i  + i\overline C \Pi _0 } \right).
\label{nueva}
\end{equation}
Hence we see that the corresponding transformations generated by
$\Omega^{\bot}$ are:
\begin{equation}
\delta ^ \bot  A_i  = -\varepsilon \partial _i \frac{1}{\nabla^2}
{\cal P}  , \label{bran1}
\end{equation}
\begin{equation}
\delta ^ \bot A_0  = i\varepsilon\overline C, \label{bran2}
\end{equation}
\begin{equation}
\delta ^ \bot \Pi_{\mu} = 0, \label{bran3}
\end{equation}
\begin{equation}
\delta ^ \bot C  =  - \varepsilon \frac{1}{\nabla^2}{\partial_i
\Pi^i }, \label{bran4}
\end{equation}
\begin{equation}
\delta ^ \bot {\overline C}=0, \label{bran5}
\end{equation}
\begin{equation}
\delta ^ \bot {\cal P}=0, \label{bran6}
\end{equation}
\begin{equation}
\delta ^ \bot  \overline {\cal P}  = -i\varepsilon \Pi _0 .
\label{bran7}
\end{equation}
Thus it follows that on integration over the momenta the
gauge-fixing term remains invariant under the transformation
generated by $\Omega^{\bot}$, that is, $\delta ^ \bot
(\partial_{\mu}A^{\mu})= 0 $. The above expressions coincide with
the Lavelle and McMullan's result \cite{Lavelle}. However, these
symmetry transformations turn out to be nonlocal. The preceding
analysis opens up the way to a stimulating discussion of how the
so-called dual BRST symmetry appears. This is precisely the task
that we shall carry out in the next section.

\section{Dual BRST symmetry}
As already mentioned, our immediate objective is to implement the
above general considerations to the two-dimensional case. With
this in mind, we start by considering
\begin{equation}
{\cal L} =  - \frac{1}{4}F_{\mu \nu } F^{\mu \nu }, \label{dual1}
\end{equation}
in two dimensions of spacetime. Just as for the four-dimensional
case, the canonical action is
\begin{equation}
S = \int {dx} \left( {\stackrel {\bf{\cdot}}{A_1} \Pi ^1  - H_0 -
\lambda \varphi } \right), \label{dual2}
\end{equation}
where $\Pi ^1$ is the momenta conjugate to $A_1$. Here it is
important to realize that the corresponding canonical Hamiltonian
is now
\begin{equation}
H_0  = \int d x\left( { - \frac{1}{2}\Pi _1 \Pi ^1 + \Pi _1
\partial ^1 A_0 } \right), \label{dual3}
\end{equation}
The constraint structure for the gauge field naturally remains
identical to the previous case ( See Eq. (\ref{gauss}) ). Thus it
follows that
\begin{equation}
\varphi  = \partial _1 \Pi ^1  = 0. \label{gauss2}
\end{equation}
Again we find that the effective action can be written as
\begin{equation}
S_{eff}  = \int {d x\left( {\Pi ^1  \stackrel {\bf{\cdot}}{A_1 } +
{\cal P}\stackrel {\bf{\cdot}}{C}+{\overline {\cal P}} \stackrel
{\bf{\cdot}}{\overline C} + \Pi^0 \stackrel {\bf{\cdot}}{A_0} -
H_0 - \left[ {\Omega ,\Psi } \right]} \right)}. \label{dual4}
\end{equation}
As in the preceding section, the BRST generator reduces to
\begin{equation}
\Omega  = \int d x\left( {C\partial _1 \Pi ^1  - i{\overline {\cal
P}}\Pi _0} \right). \label{carga}
\end{equation}
We can now write the corresponding transformations generated by
$\Omega$, that is,
\begin{equation}
\delta A_1  = - \varepsilon \partial _1 C , \label{dual5}
\end{equation}
\begin{equation}
\delta A_0  = -i\varepsilon {\overline {\cal P}} , \label{dual6}
\end{equation}
\begin{equation}
\delta \Pi_1 = 0 , \label{dual7}
\end{equation}
\begin{equation}
\delta \Pi _0 = 0 , \label{dual8}
\end{equation}
\begin{equation}
\delta C = 0 , \label{dual9}
\end{equation}
\begin{equation}
\delta {\overline C} =  i\varepsilon  \Pi _0, \label{dual10}
\end{equation}
\begin{equation}
\delta {\overline {\cal P}} = 0, \label{dual11}
\end{equation}
\begin{equation}
\delta {\cal P} = -\varepsilon \partial _1 {\Pi ^1}.
\label{dual12}
\end{equation}
Following our procedure we now calculate the effective action
(\ref{dual4}). As in the previous section, we choose the
gauge-fixing function in the form
\begin{equation}
\Psi  = \int d x \left( {{\cal P}A_0  - i{\overline C}\left(
{\partial _1 A ^1 - \frac{\xi }{2}\Pi _0 } \right)} \right).
\label{dual13}
\end{equation}
In the present case, the effective action is found to be
\begin{equation}
S_{eff}  = \int {d^2 x\left( { - \frac{1}{4}F_{\mu \nu } F^{\mu
\nu }  - \frac{1}{{2\xi }}\left( {\partial _\mu  A^\mu  }
\right)^2  + i\overline C \partial _\mu  \partial ^\mu  C}
\right)}. \label{dual14}
\end{equation}

It has been recently claimed \cite{Malik} that the effective
action is also invariant under local variations:
\begin{equation}
\delta _D A_\mu   =  - \eta \varepsilon _{\mu \nu } \partial ^\nu
\overline C ,\label{mali1}
\end{equation}
\begin{equation}
\delta _D \overline C  = 0, \label{mali2}
\end{equation}
\begin{equation}
\delta _D C =  - i\eta {\cal B}, \label{mali3}
\end{equation}
where $ \cal B $ is an auxiliary field. Accordingly, we have that
\begin{equation}
\delta _D \left( {\partial _\mu  A^\mu  } \right) = 0.
\label{mali4}
\end{equation}
At this point it is reasonable to ask how the transformations
(\ref{mali1}- \ref{mali3}) are related with the ones (\ref{dual5}-
\ref{dual12}). In view of this situation and on the basis of the
discussion in the previous section, we now proceed to perform a
canonical transformation in the ghost sector. In that case, we
propose the following canonical transformation
\begin{equation}
C^\prime   = i\frac{{\cal P}}{{\partial _1 }}, \label{cano1}
\end{equation}
\begin{equation}
P^\prime   =  - i\partial _1 C, \label{cano2}
\end{equation}
\begin{equation}
\overline C ^\prime   =  - i\frac{{\overline {\cal P} }}{{\partial
^1 }}, \label{cano3}
\end{equation}
\begin{equation}
\overline P ^\iota   = i\partial ^1 \overline C. \label{cano4}
\end{equation}
As before, we keep the notation $\Omega^{\bot}$ for the charge
which results from a canonical transformation. Thus, the new
charge may be rewritten as
\begin{equation}
\Omega ^ \bot   = \int {dx\left( {i\frac{{\cal P}}{{\partial _1
}}\left( {\partial _1 \Pi ^1 } \right) - \partial ^1 \overline C
\Pi _0 } \right)}. \label{genera}
\end{equation}
It is now once again straightforward to work out the
transformations generated by (\ref{genera}). They are
\begin{equation}
\delta ^ \bot  A_1  = i\varepsilon {\cal {\cal P}}, \label{nue1}
\end{equation}
\begin{equation}
\delta ^ \bot  A_0  =  - \varepsilon \partial ^1 \overline C,
\label{nue2}
\end{equation}
\begin{equation}
\delta ^ \bot \Pi _\mu   = 0 , \label{nue3}
\end{equation}
\begin{equation}
\delta ^ \bot \overline C  = 0, \label{nue4}
\end{equation}
\begin{equation}
\delta ^ \bot C = - i \varepsilon {\Pi ^1}, \label{nue5}
\end{equation}
\begin{equation}
\delta ^ \bot {\cal P} = 0, \label{nue6}
\end{equation}
\begin{equation}
\delta ^ \bot \overline {\cal P} = - \varepsilon {\partial ^1}
\Pi_0. \label{nue7}
\end{equation}
One immediately sees that, on integration over the momenta, the
above transformations (\ref{nue1}-\ref{nue7}) reduce to the ones
found in \cite{Malik}. It is important to realize that, after
integration over the momenta, the new transformations yield
$\delta ^ \bot (\partial_{\mu}A^{\mu})=0$ off shell. It is
worthwhile sketching at this point our procedure. As mentioned
before, in the extended phase space we have $\delta \left(
{\partial _\mu A^\mu  } \right) =  - \varepsilon \left( {i\mathop
{\overline {\cal P} }\limits^ \cdot   + \partial_{1}\partial^{1}
C} \right)$, which in the configuration space reads $\delta \left(
{\partial _\mu A^\mu } \right) =\varepsilon
\partial_{\mu}\partial^{\mu}C$, but this is just the classical
equation of motion of $C$. From our above analysis, we see that
the proposed canonical transformation makes a change of the ghost
equations, that is, $ \delta ^ \bot \left( {\partial _\mu  A^\mu
} \right) =  - \varepsilon \partial ^1 \left( {\mathop {\overline
C }\limits^ \cdot   - i{\cal P}} \right) $ which after integration
over the momenta gives zero, turning the variation of the
gauge-fixing term null on shell to null off shell. Since the
canonical transformation has been carried out in the ghost sector,
all the basics processes that can be explained by the old
effective action, should likewise be obtained from the new
effective action. It is satisfying to notice the simplicity and
directness of this derivation, which is manifestly
gauge-independent.

\section{ACKNOWLEDGMENTS}

The author would like to thank I. Schmidt for his support, and
J.J. Yang for reading the manuscript. Work supported in part by
Fondecyt (Chile) grant 1000710, and by a C\'atedra Presidencial
(Chile).

\end{document}